\documentclass[superscriptaddress,twocolumn,nofootinbib]{revtex4}
\usepackage{graphicx}
\usepackage{color}
\usepackage{enumerate}
\usepackage{amssymb}
\usepackage{bbm}
\usepackage{dsfont}
\def\be{\begin{equation}} 
\def\ee{\end{equation}}

\def\inn{{\rm in}}

\begin{document}

\title{
Configuration-interaction approach to nuclear fission 
}

\author{G.F. Bertsch}
\affiliation{ 
Department of Physics and Institute of Nuclear Theory, Box 351560, 
University of Washington, Seattle, Washington 98915, USA}

\author{K. Hagino}
\affiliation{ 
Department of Physics, Kyoto University, Kyoto 606-8502,  Japan} 

\begin{abstract}
We propose a configuration-interaction (CI) representation to calculate
induced nuclear fission with explicit inclusion of nucleon-nucleon
interactions in the Hamiltonian.  The
framework is designed for easy modeling of schematic interactions
but still permits a straightforward extension to realistic ones. 
As a first application, the model is applied to branching ratios between
fission and capture in the  decay modes of excited fissile nuclei.  
The ratios are compared with the Bohr-Wheeler transition-state theory
to explore its domain of validity.  The Bohr-Wheeler theory 
assumes that the rates are insensitive to the final-state scission
dynamics; the insensitivity is rather easily achieved in the CI parameterizations.
The CI modeling is also capable of reproducing the branching ratios of
the transition-state hypothesis which  is one of the key ingredients
in the present-day theory of induced fission.
\end{abstract}

\maketitle

{\it Introduction.} The theory of induced fission 
is one of the most challenging subjects in many-fermion quantum dynamics.
In a recent review \cite{be20b} of future directions in fission theory, the 
authors omitted the topic ``because there has been virtually no coherent 
microscopic theory addressing this question up to now." 

In this Letter we propose a microscopic  approach based on many-body Hamiltonians in the
configuration-interaction (CI) framework.  The idea is not new \cite{do89},
but the methodology has yet to be applied in practice\footnote{There has
been earlier work calculating the dynamics by a diffusion equation with a
microscopic treatment of the diffusion coefficient \cite{BBB92,CB92}.}. 
Before realistic
calculations can be contemplated, it is useful to consider simplified
models in the many-particle framework that are extensible to the realistic
domain \cite{be20,ha20,ha20-2}. 
Such models may validate the phenomenological
approaches that have been with us since the beginnings of fission theory,
or it may suggest modifications to them.
The focus here is on how the system crosses the fission barrier;
key observables are the excitation function for fission cross sections 
and the branching ratio between fission and other decay channels.
The model proposed below 
incorporates microscopic mechanisms to propagate the systems from the
initial  ground-state shape to a region beyond the fission 
barriers(s).

{\it Hamiltonian. } 
We build the Hamiltonian on a set of
reference states $Q$, each such state generating a spectrum of
quasiparticle excitations which we call a $Q$-block.  The $Q$-blocks
are ordered by deformation $Q_2$.
The Hamiltonian is constructed from these elements as
\be
\hat H = \sum_Q \left( E_{\rm gs}(Q) + \hat H_{\rm qp}(Q) + \hat H_{v}(Q)
 + f_{\rm od}\sum_{Q'=Q\pm1}   \hat H_{v}^{\rm od}\right).
\label{H} 
\ee
Here $E_{\rm gs}$ is the energy of the reference state,
calculated by constrained Hartree-Fock or density functional
theory\footnote{We note that theory based on Hamiltonian
interactions together with orbitals from DFT has been
successfully applied elsewhere \cite{sa16}.} (DFT).  In the model
below we choose appropriate sets of energies $E_{\rm gs}(Q)$
to explore various limits of the theory.  The
circumflexes denote terms containing Fock-space operators
acting within a $Q$-block or between orbitals in adjacent
$Q$-blocks.  We detail them below.

{\it Constructing the configurations.}
The configuration space is built in the usual way, defining
configurations as Slater determinants of nucleon orbitals.  The
orbitals are envisioned as eigenstates of an axially
deformed single-particle potential.  Ultimately their properties
would be determined by the density functional theory, but for
modeling purposes we found it convenient to assume a
uniform spectrum of orbital energies with the same spacing $d$
for protons and neutrons.  The ladder of orbital states 
extends infinitely in both directions above and below the Fermi
surface. 
The operator for the quasiparticle
excitation energy $E_{qp} $ is given by
\be
\hat H_{\rm qp} = d \sum_{\alpha:~n_\alpha > 0}  n_\alpha \hat a^\dagger_\alpha
\hat a_\alpha + d  \sum_{\alpha:~n_\alpha \le 0}  n_\alpha \hat a_\alpha
\hat a^\dagger_\alpha.
\ee
The label $\alpha$ includes all quantum numbers associated with the
orbital, $ \alpha = (Q,n,K,t)$.  Here $n$ indexes the orbital
position in the ladder, with $n=0$ corresponding
to the Fermi level, and $K$
is its angular momentum about the
symmetry axis.  To keep the model as transparent as possible, we
restrict $K$ to $\pm 1/2$.  The label $t$ distinguishes
neutrons (n) and protons (p).  

\begin{figure}
\includegraphics[clip, trim=0.0cm 0.0cm 0.0cm 0.0cm,width=1.0\columnwidth]{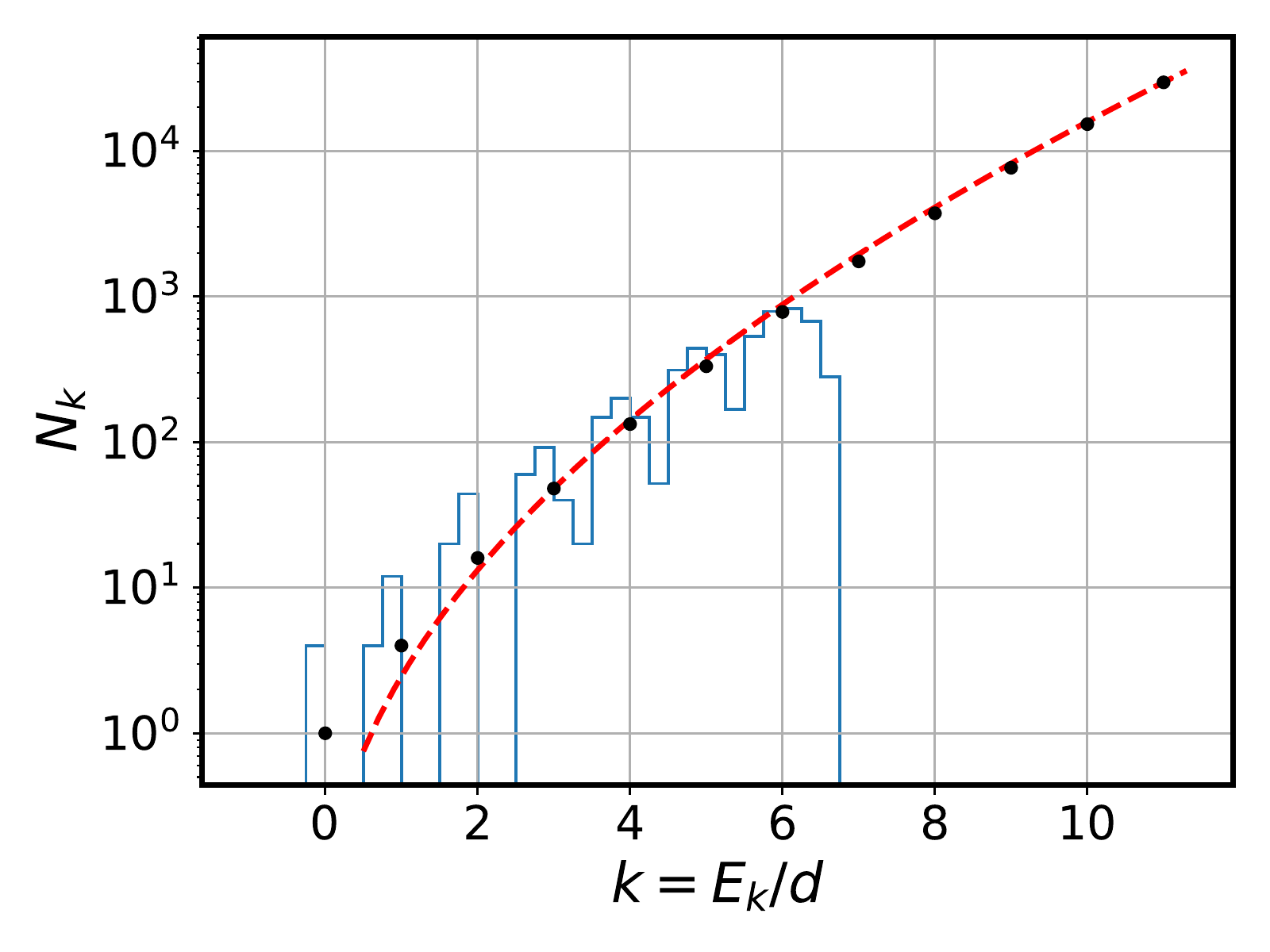}
\caption{
\label{spectrum}
Spectrum of many-body configurations in the uniform model. 
Here, the total $K$ quantum number of the system is restricted to $K=0$. 
$N_k$ denotes the number of configurations at the excitation 
energy $E^*=kd$. The filled circles and the histogram 
show the non-interacting and the interacting spectra, respectively. 
The dotted red curve shows a fit to
the functional form $\log(N_k) \sim \sqrt{k}$.
}
\end{figure}

The orbital excitation energies of many-particle
configurations are integral multiples of $d$, $E_{\rm qp} = E_k = k\, d$.  
As a function of
$k$, the multiplicity of configurations 
having $\sum K=0$ 
is 
$N_k = (1,4,16,48,133,332,784,...)$ for $k=(0,1,2,3,4,5,6,\cdots)$.  
The  spectrum up to $k=11$ is shown in Fig. \ref{spectrum}.
\begin{figure}
\includegraphics[clip, trim=0.0cm 0cm 0.0cm 0.0cm,
width=0.8\columnwidth]{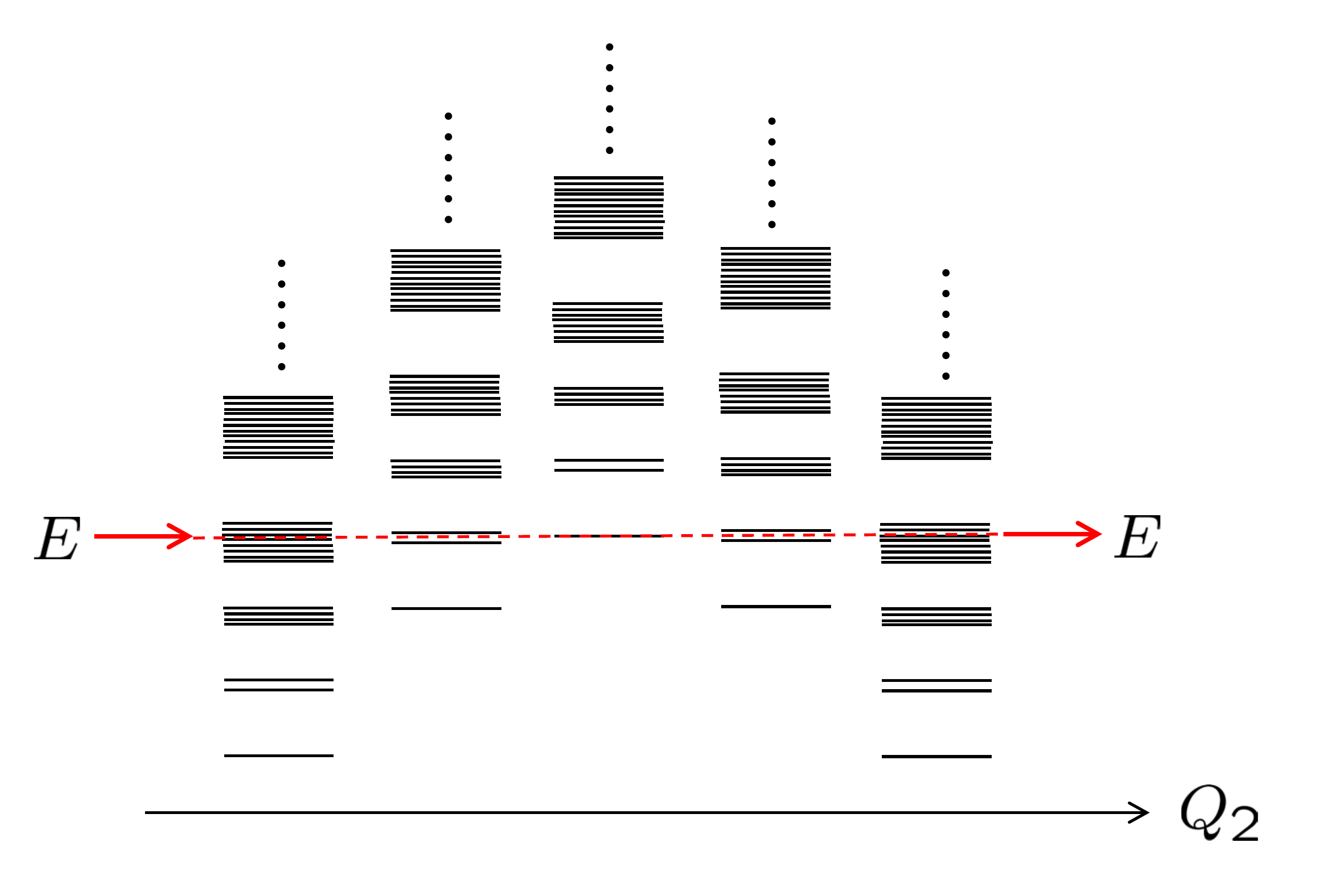}
\caption{
\label{Q-blocks}  The vertical towers of levels are the
$Q$-blocks in the basis of configurations. 
They are ordered by deformation $Q_2$. $E$ is the incident energy 
for a fission reaction. 
}
\end{figure}
Its functional form agrees well with the leading behavior of the Fermi-gas
level density formula \cite{bm1} as described in the figure caption. The
histogram shows the level density after including $\hat H_v$ in the
Hamiltonian of the $Q$-block.    

The parameter $d$ will be left unspecified below.  It can vary greatly due
to shell effects, but in the actinide nuclei it is in the 
range 
$0.4-0.5$ MeV \cite{Supp}.  The levels at the neutron emission threshold
would then correspond to $k\approx 13-16 $ in excitation energy and  somewhat
higher $k\approx 17$ in level density.  
What we have described here is a single-reference basis of configurations.
Fission requires large-amplitude shape changes, which cannot be reasonably
treated in a single-reference basis.  As a minimum, one needs to
extend the space by including as reference states the local DFT minima
across the saddle point of the barrier \cite{barranco90}.  In fact there are many such
minima along typical fission paths \cite{be19a}.  In our model
we organize the reference configurations as a chain along a path
of increasing deformations, as depicted in Fig. \ref{Q-blocks}.  One 
complication at this point is that the resulting basis may not be
orthogonal. We shall come back to this point later.
 
{\it Model nucleon-nucleon Hamiltonians.}
In the quasiparticle representation, there are three kinds of 
two-particle interaction.
The interactions that are diagonal in quasiparticle occupations factors
are taken into account in $E_{\rm gs}$, the ground state energy of the reference
configuration.  The interactions changing the orbital of one of the
nucleons do not contribute  in the reference configuration if it is a 
stationary state of the DFT; otherwise it induces a diabatic transformation
of the configuration.  Such transformations are generally unfavored on 
energetic grounds \cite{no83} and they are omitted in the
present model.
We are left with the interactions that change the orbitals of both
particles.  In this Letter we deal only with the neutron-proton
interaction.   The pairing interaction between identical particles is 
certainly important as well; in fact, it is likely to be more important
in non-diabatic collective dynamics \cite{ha20-2,ro14}.  However, it is
also important to assess the effects of the residual neutron-proton
interaction \cite{BBB92}, and this has not be done until now.

We write the interaction Hamiltonian as
\be
\hat H_{v} = v_{\rm np} {\sum}'
 r\,
\hat a^\dagger_{\alpha_1}\hat a^\dagger_{\alpha_2}\hat a_{\alpha_3}
\hat a_{\alpha_4}.  
\label{v-eq}
\ee
where the parameter $v_{\rm np}$ is the strength of the interaction and $r$ is
a random variable from a Gaussian ensemble of unit variance. The summation is
over the four $\alpha$ indices restricted 
to a fixed $Q$ in $\hat H_v$ and to neighboring $Q$-blocks 
\cite{BBB92,CB92,be20,ha20,ha20-2,barranco90} 
in $\hat
H^{\rm od}_v$.
Also, the sum is restricted to $\alpha$ sets satisfying
$K_1+K_2 = K_3 + K_4$. 
The assumption that the neutron-proton interaction is Gaussian distributed
is certainly not justified for the low-energy states in a $Q$-block where
collective excitations can be built up. However,  high in the spectrum 
where only the
overall interaction strength is important
the
mixing approaches the random matrix limit.  The strength can be
determined by sampling with more realistic interactions that could
range in sophistication from simple contact interactions or 
separable interactions to those used in present-day shell model
Hamiltonians.  
The strength of neutron-proton contact interactions for shell
model Hamiltonians is typically in the range $250-500$ MeV-fm$^3$ 
\cite{BBB92,yoshida13}.  The
corresponding strength in actinides \cite{Supp}
 for our parameterization is $0.05 d-0.1 d$; 
we take $v_{\rm np} = 0.05d$ in most of the examples below.

The configurations may be characterized by the number of quasiparticles
as well as by the energy index $k$. Each 
$k\ne 0$  subblock contains 
configurations  
going from two quasiparticles to the maximum energetically allowed.  
The subblocks  are
all connected by the residual interaction, although the matrices 
connecting them are sparse.  
For example,
the $N_6\times N_6$  matrix has an off-diagonal filling of
5\%, while the $N_6\times 1$ matrix connecting the $k=6$
subblock to the ground state is 27\% filled.  

In the presence of the residual interaction, the eigenstates of a
shell-model configuration space are found to approach the random matrix
limit of the Gaussian Orthogonal Ensemble (GOE) when the rms interaction
strength is larger than the level spacing between 
configurations \cite{gu89,wei09}.    For reaction theory,
the most important GOE characteristic is the Porter-Thomas distribution
of decay widths,  requiring a nearly Gaussian distribution
of configuration amplitudes in the eigenstates.
The eigenstates of large-dimension
$k$-subblocks do in fact acquire the properties of the GOE, 
even though the sparseness of the interaction matrix works
against a complete mixing of the configurations \cite{Supp}. 
More realistic
model that do not permit the $k$ grouping will still approach the
GOE at high excitation energy.  However, it should be mentioned that
a numerical study \cite{br84,zel96} of a light-nucleus spectrum did not confirm the
above stated criterion for GOE behavior.

The interaction between $Q$-blocks is responsible for shape
changes \cite{BBB92} and is thus crucial to the modeling.  It is clear that
the interaction is somewhat suppressed due to the
imperfect overlap of orbitals built on different mean-field reference
states.  Another complication is that the configurations in 
different $Q$-blocks will not be
orthogonal unless special measures are taken, e.g., restriction by
$K$-partitioning \cite{be17}.  These problems have long been dealt with
in other areas of physics \cite{lo55,re11,ro20} and can be treated in
nuclear physics in the same way.  For our model, we simply 
parameterize the effects by the attenuation factor $f_{\rm od}$ in Eq. (1).

{\it Reaction theory.}
Induced fission is in the domain of reaction theory: an external
probe, typically a neutron, excites the nucleus leading to its
decay by fission.  A number of reaction-theoretic formalisms 
are available for treating CI Hamiltonians.  We mention in particular\footnote{Early 
studies also made use of the $R$-matrix theory \cite{la58,thompson09,KTW15}.  However, it
requires unphysical boundary conditions that are difficult to
implement.}  the 
$K$-matrix formalism \cite{be20,KTW15,BK17,al20,al21} 
and the $S$-matrix formalism \cite{KTW15,thompson09}.
The key quantity is the transmission coefficient from an incoming
channel to the decay channels of interest, 
\be
T_{\inn,C} = \sum_{j \in C} |S_{\inn,j}|^2.
\ee
Here $C$ is the set of quantum mechanical channels associated with
the type of reaction.  For neutron-induced reactions on heavy nuclei, 
it could be inelastic scattering, capture, or fission.  The relevant
$S$-matrix quantities may be calculated as \cite{da01,al21}
\be
|S_{j,j'}|^2  =  \sum_{\mu,\mu'}\Gamma_{j,\mu}
 |( \tilde H - E)^{-1}|^2_{\mu,\mu'} \Gamma_{j',\mu'}. 
\label{S2}
\ee
Here  $\Gamma_{i,\mu}$ is the decay width of the state $\mu$
into the channel $i$.  Note that the CI Hamiltonian $H$ is modified by 
including
of the coupling to the channels.  We assume in the model that each channel couples
to a single internal configuration, and we neglect dispersive effects.
The modified Hamiltonian $\tilde H$ then reads,
\be
\tilde H_{\mu,\mu'} = H_{\mu,\mu'} - i \delta_{\mu,\mu'} \sum_j
\Gamma_{j,\mu}/2.
\ee
The main observable we are interested in is the branching ratio between
fission and capture.  We define it as
\be
B_{\rm cap,fis} = \frac{\int d E \,T_{\inn,{\rm fis}}}
{\int d E \,T_{\inn,{\rm cap}}}.
\label{B}
\ee
The range of integration is the same for numerator and denominator
and in practice would be determined by experimental considerations.  For
simplicity, we assume that the entrance channel width is small
compared to the decay widths, in which case it cancels out of
Eq. (\ref{B}).
For a
typical example the experimental quantities are $\Gamma_{\rm cap}
\approx 0.04 $eV and $B_{\rm cap,fis} \approx 3$ \cite{BK17}.  

{\it Results. }  
We can now set the parameters to simulate the
branching between capture and fission processes.
To this end, we consider chains of three or more $Q$-blocks; the first
represents the spectrum built on the ground
state and the last has the doorway state to fission channels.
Imaginary energies $-i\Gamma_{\rm cap}/2$ and $-i\Gamma_{\rm fis}/2$ are
added to the  blocks in $\tilde H$ to account for the
decay widths \cite{be20}. 

\begin{figure}
\includegraphics[width=0.8\columnwidth]{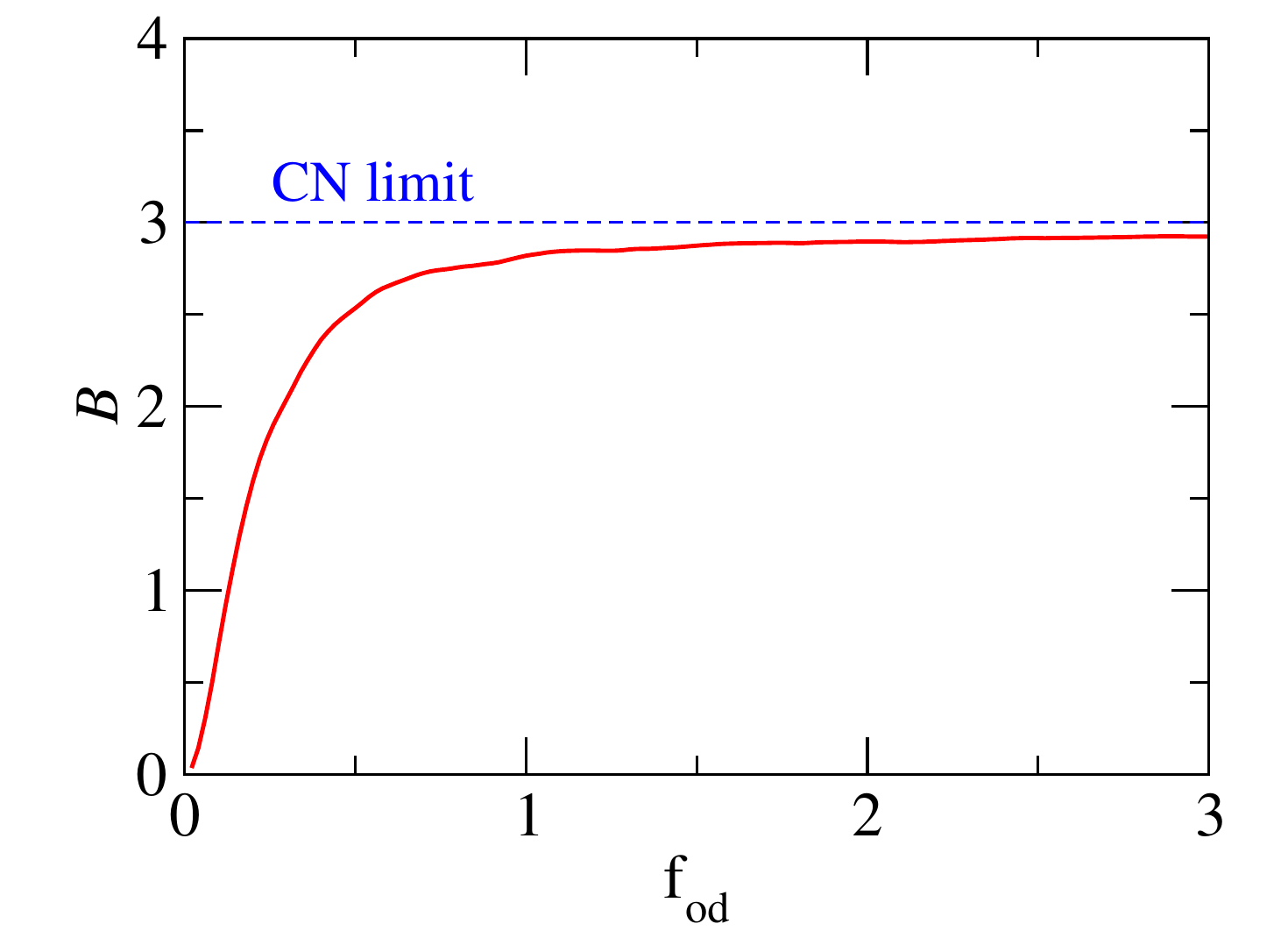}
\caption{
The branching ratio as a function of the attenuation factor $f_{\rm od}$ 
in	 the interaction between $Q$-blocks. 
The model space is $(k=0-4)\times 3$  and the reference-state energies
are
$E_{\rm gs} = (0.0,0.0,0.0)$. 
Other parameters are: $v_{\rm np} = 0.05d$, 
$\Gamma_{\rm cap}= 0.001d$, and $\Gamma_{\rm fis}= 0.003d$. 
The energy average in Eq. (\ref{B}) is taken in the range of 
$3.5d \leq E \leq 4.5d$. 
The dashed line denotes the branching ratio in the compound 
nucleus limit, 
$B_{\rm cap, fis} =\Gamma_{\rm fis}/\Gamma_{\rm cap}=3.0$. 
}
\label{cn}
\end{figure}

As a warm-up, we find conditions on $f_{\rm od}$ that justify the
compound-nucleus (CN) hypothesis that the relative decay rates
are proportional to the decay widths in $\tilde H$.  The model has 
three identical $Q$-blocks
composed of $k\le 4$ subblocks.  The calculated branching ratios 
are shown in Fig. \ref{cn}.  Note that the stochastic treatment of
the interaction in Eq. (\ref{v-eq}) produces a distribution of 
ratios; only one of them is shown in the figure.
For $f_{\rm od}\gtrsim 1$, the branching
ratio  is consistent with the formula 
$B_{\rm cap, fis} =\Gamma_{\rm fis}/\Gamma_{\rm cap}$, confirming the CN limit.  Even
with $f_{\rm od} = 0.5$, the branching ratio is only $15$\% less than
the CN limit.

We next impose a barrier and examine 
the fundamental 
assumption of present-day theory of induced fission, namely that
near the barrier top the decay rate is by the 
Bohr-Wheeler (BW) transition state formula \cite{BW39,B91}
\be
\Gamma_{\rm BW} = \frac{1}{2 \pi \rho} \sum_i T_i.
\label{BW}
\ee 
Here $i$ are states on the barrier top, $T_i$ are
transmission coefficients across the barrier, and $\rho$ is the 
level density of the compound nucleus (i.e., the first $Q$-block) at the 
given excitation energy. Notice that the BW formula does not
depend on the fission widths $\Gamma_{\rm fis}$, unlike Eq.
(\ref{S2}).
\begin{figure}
\includegraphics[clip, trim=0.0cm 0cm 0.0cm 
0.0cm,width=0.8\columnwidth]{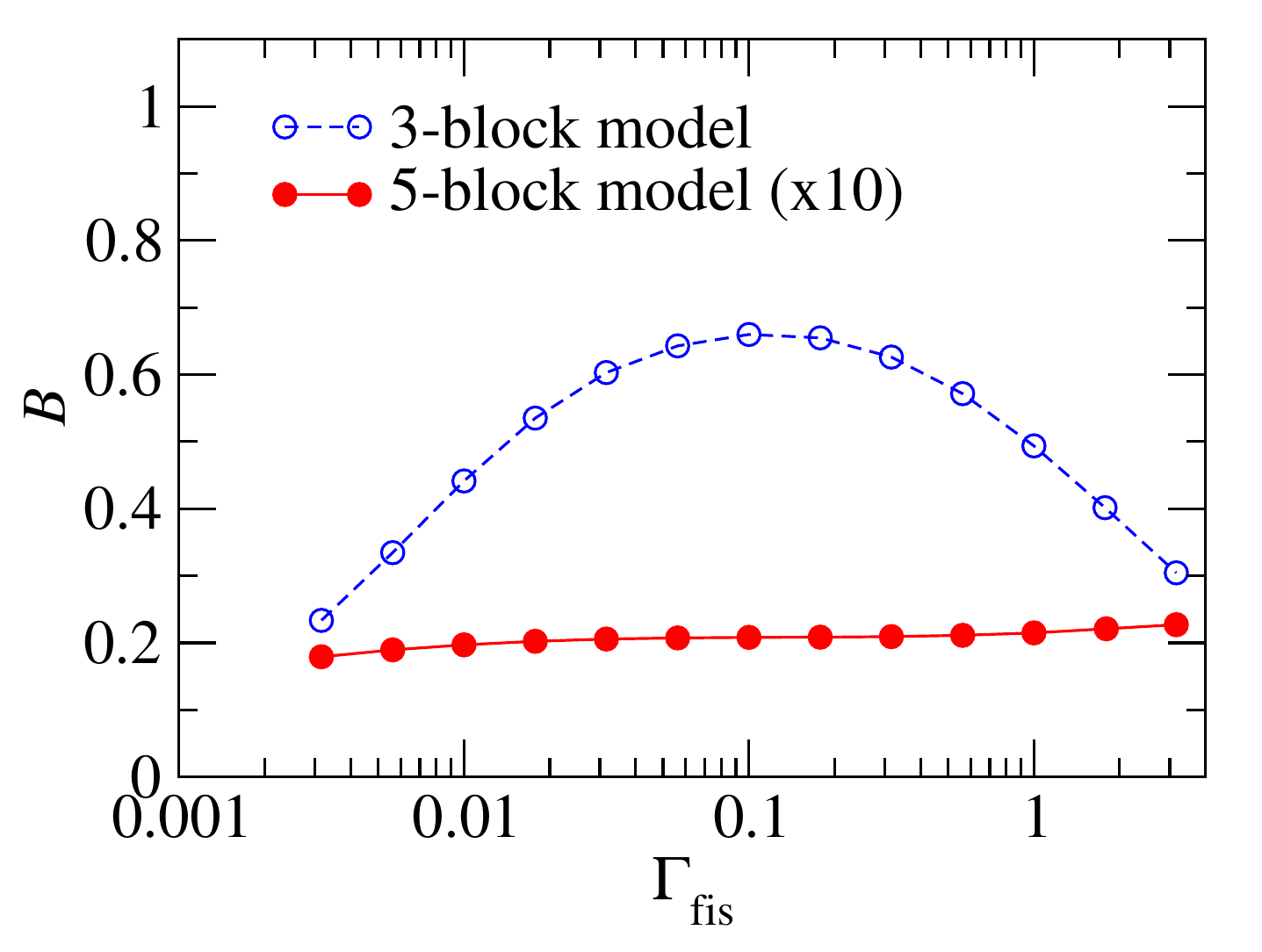}
\caption{
\label{sensitivity}
Sensitivity of the branching ratio to the exit channel
widths.  
The blue open circles are for the model space 
of $(k=0-6)\times 3$  with
$E_{\rm gs} = (0,6d,0)$, while the red filled circles are 
for $(k=0-6)\times5$ with a parabolic barrier $E_{\rm gs} = (0.0,4.5,6.0,4.5,0.0)$.
Other parameters are: $v_{\rm np} = 0.05 d$, $f_{\rm od} = 1.0$, and 
$\Gamma_{\rm cap}= 0.01d$. 
The incoming channel is assumed to be one of the configurations at $E^*=6d$ 
in the first block, and 
the energy average in Eq. (\ref{B}) is taken in the range of 
$5.5d \leq E \leq 6.5d$. 
}
\end{figure}

First consider the model space of 3 identical $Q$-blocks
composed of $k\le6$ subblocks, with $E_{\rm gs} = (0,6d,0)$ to
make a barrier at the middle block.
As may be seen in 
Fig. \ref{sensitivity}, this model fails the
first assumption:  the derived branching ratio is sensitive to
the fission decay width, $\Gamma_{\rm fis}$.  The reason is that there are 
many virtual transitions possible through the higher levels
in the barrier-top $Q$-block.  Because the effective number of partially
open channels is large, the communication between
the end $Q$-blocks remains strong.

We found two ways to greatly diminish the dependence on
$\Gamma_{\rm fis}$ in our model.
The first way is to increase the chain of $Q$-blocks on
the barrier.  Then the path across the barrier requires
multiple virtual transitions, resulting in a much stronger
suppression factor. This may be seen in Fig.  \ref{sensitivity} 
for the 5-block case. 

\begin{figure}
\includegraphics[width=0.8\columnwidth]{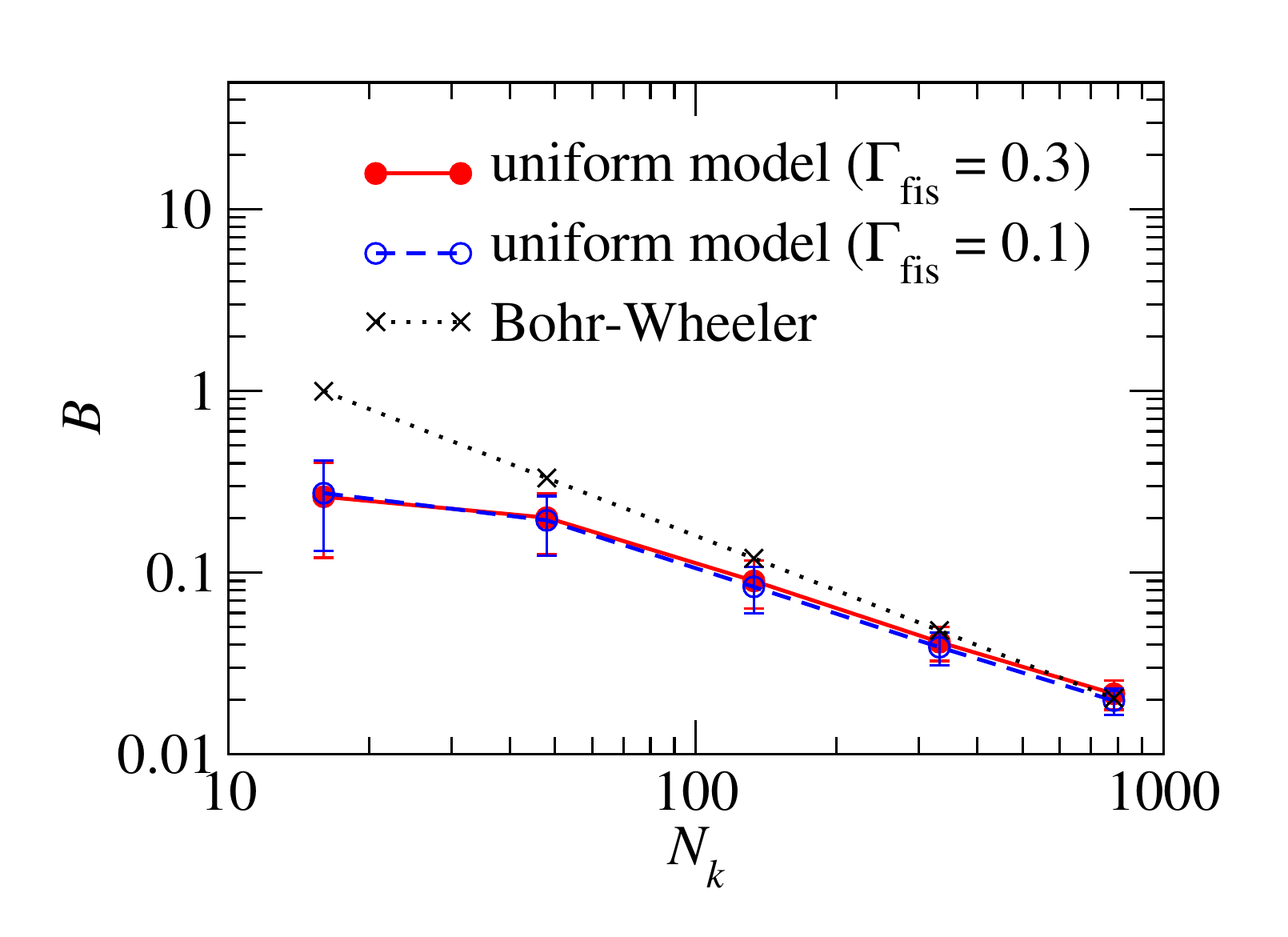}
\caption{
\label{BW-fig}
Branching ratios in $(k)/(0)/(k)$ configuration spaces in which the
barrier $Q$-block is a single state degenerate with the other 
$Q$-blocks.  Other parameters are $v_{\rm np}=0.05 d,~f_{\rm od}=1$ 
and $\Gamma_{\rm cap}=0.01d$. 
The filled circles and the open circles are obtained with 
$\Gamma_{\rm fis}=0.3d$ and $0.1d$, respectively. The uncertainties are estimated 
with 100 different random number sets in 
the Hamiltonian matrix. 
The filled squares shows the
predicted values from a schematic transition-state formula (see text for details).
}
\end{figure}

The other way is to eliminate  the virtual transitions at the
barrier by cutting off the spectrum of the middle block.  Fig. 
\ref{BW-fig} shows the results with 3 Q-blocks.  The first and
the last blocks are defined as usual in the $N_k$ space.
The middle block has only a $k=0$ configuration,
shifted in energy to $E_{\rm qp} = k\, d$.  The filled circles and
the open circles show the branching ratios with $\Gamma_{\rm
fis}$ = $0.3d$  and $0.1d$, respectively.  There is hardly any
difference between the two curves.

One can make a 
crude approximation to the Bohr-Wheeler transition
state formula Eq. (\ref{BW}) within the framework of the
model.
The branching ratio for a single barrier
state with a transmission factor $T_t = 1$ 
reads $B_{\rm cap, fis} \sim 1/(2\pi \rho\Gamma_{\rm cap})$. 
The spread of the $k$ subblock with the given
parameters is about $d$, resulting in a level density
$\rho = N_k /d $.  This estimate gives the reasonable agreement
shown by the filled squares in the Figure; it appears that
the internal transmission factor for large spaces approaches
$T=1$.  However, the comparison should not be considered 
quantitative because the level density of the first $Q$-block
is not constant over
the energy window accessed by the state in the middle.

{\it Summary and Outlook.}  
The model presented here for induced fission in a  
CI representation appears to be 
sufficiently detailed to examine the validity of transition-state
theory in a microscopic framework.  Depending on the interaction
and the deformation-dependent configuration space,  one achieves
conditions in which branching ratios depend largely on barrier-top
dynamics and are insensitive to properties closer to the scission
point.  
The insensitive property is 
one of the main assumptions in the well-known Bohr-Wheeler formula 
for induced fission, but up to now it had no microscopic justification.

Whether the transition-state hypothesis is valid under realistic
Hamiltonians remains to be seen and will require a large
computational effort to answer.  In the near term, the model can
be applied in a number of different ways.  We plan to study the
barrier transmission factor $T_i$ as a function of barrier
height to test another basic assumption in present-day theory,
namely treating them by the Hill-Wheeler
formula \cite[p.~1140]{hi53}.  It also appears quite
straightforward to include a pairing interaction in the
Hamiltonian.  This would allow one to explore for the first time
the competition between the two kinds of interaction in
barrier-crossing dynamics.

\medskip

This work was supported in part by 
JSPS KAKENHI Grant Number JP19K03861.

\end{document}


\noindent{\bf Contents}\\
1. Estimates of the physical parameters\\
2. Compound nucleus limit\\
3. Codes\\

\noindent{\bf Estimation of physical parameters}\\

\noindent{\it Orbital energy spacing $d$.}
The single-particle level spacing in the uniform model
$d$
sets the energy scale for the model and does not
play any explicit role in the model.  However, it is required to
determine other energy parameters which are expressed
in units of $d$.  Several estimates of $d$ for \uf~ 
are given in Table I.  The first is based on orbital energies
in a deformed Woods-Saxon potential with the parameters given 
in Ref. \cite{bm1}; see Table II for the
calculated orbital energies.
\begin{table}[htb] 
\begin{center} 
\begin{tabular}{|c|c|} 
\hline
$d$ (MeV) &  Source \\
\hline
0.45  &  Woods-Saxon well \\
0.51 &  FRLDM  \cite{mo95} \\
0.33 &  FGM  \cite{ko08} \\
\hline
\end{tabular} 
\caption{Estimated orbital level spacing in \uf.  The first
two are from potential models and the last extracted from
the Fermi gas formula and measured level densities.
}
\label{orbital-d} 
\end{center} 
\end{table} 
\begin{table}[htb] 
\begin{center} 
\begin{tabular}{|ccc|ccc|} 
\hline 
\multicolumn{3}{|c|}{protons} & \multicolumn{3}{c|}{neutrons}\\
\hline
$2K$ & $\pi$ & $\varepsilon_{K\pi}/d$ & $2K$ & $\pi$ & $\varepsilon_{K\pi}/d$ \\
\hline
3 & $-$1 & $-$3.39 & 5 & $-$1  & $-$4.15 \\
5 & $-$1 & $-$3.80& 1 & $-$1 & $-$4.25 \\
5 &  1& $-$4.93 & 7 & $-$1 & $-$4.40 \\
\multicolumn{3}{|c|}{- - - - - -} & \multicolumn{3}{c|}{- - - - - -}\\
1 & 1& $-$5.43 & 1 & 1 & $-$5.07  \\
9 & $-$1& $-$5.53 & 5 & 1 & $-$5.75 \\
3 &  1 & $-$5.74 & 5 & $-$1 & $-$5.82 \\
\hline
\hline\end{tabular} 
\caption{Characteristics of single-particle orbitals in a deformed
Woods-Saxon potential corresponding to \uf~ at deformation
$(\beta_2,\beta_4) = (0.274,0.168)$. Dashed line indicates the Fermi level.
}
\label{orbital-e} 
\end{center} 
\end{table} 
In
more realistic theory, the momentum dependence of the potential
tends to increase the spacing, but the coupling to many-particle
degrees of freedom decreases the spacing of the quasiparticle 
poles.  The combined effect seems to be to somewhat decrease
the spacing\footnote{We note that in the energy density functional
fit including fission data\cite{ko12} the effective mass in the single-particle
Hamiltonian was very close to  1.}.\\  

\noindent{\it Level density.} It is important to know the composition 
of the levels in the
compound nucleus to construct microscopic models that involve
those levels.  For a concrete example, consider the levels at
the neutron threshold energy $S_n = 6.5$ MeV in $^{236}$U.
The predominating configurations at this energy should be
$k$ subblocks at $k\approx S_n/d$ in the independent quasiparticle
approximation.  Another approach that is less sensitive to the
residual interaction is to estimate the total number of states
below $S_n$ and comparing it to the number obtained by summing
the $N_k$ degeneracies in the $Q$-block spectrum.  In the
\uf~ example, the combined level spacing of $J^\pi = 3^-$ and
$4^-$ is about 0.45 eV at $S_n$\cite{ca09}.  At that excitation energy
the level density is the same for even and odd parities, and it
varies with angular momentum as $2J+1$.  The inferred 
level spacing of $J^\pi = 0^+$ levels is thus about 7 eV. The
accumulative number of levels can be approximated  by
$N = \rho T$ where $T$ is the nuclear temperature, defined as
$T = d \log(\rho(E))/d E$.  A typical estimate for our example
is $T= 0.65$ MeV,  giving $N \approx 1.0\times10^8$. 
To estimate the level density in the present model, we start
with the set of quasiparticle configurations including both
parities and all $K$ values.  The resulting $k$-blocks have
multiplicities that are well fit by the formula 
\be
N_k \approx \exp(-3.23 + 4.414 k^{1/2}).
\ee 
Projection on good parity decreases this by a factor of two.  The
projection on angular momentum $J=0$ is more subtle.
The $J=0$
states are constructed by projection from $K=0$ configurations; other
configurations do not contribute.  However, there may be two distinct
configurations that project to the same $J=0$ state.  This gives
another factor of nearly two reduction in the multiplicity.  The
remaining task is to estimate the fraction of $K=0$ configurations
in the unprojected quasiparticle space.  The distribution of 
$K$ values is approximately Gaussian with a variance given by
\be
\langle K^2\rangle = \langle n_{\rm qp}\rangle \langle K^2\rangle_{\rm sp}
\ee
where $\langle n_{\rm qp}\rangle\approx 8$  is the average number of quasiparticles
in the $k$ block and $\langle K^2\rangle_{\rm sp} \approx 6$ is an
average over the orbital $K$'s near the Fermi level.  Including
these projection factors, the integrated number of levels up
to $S_n$ is achieved by including all $k$-subblocks up to $k=17$
in the entry $Q$-block. \\

\noindent{\it Neutron-proton interaction $v_{\rm np}$.}
To set the scale for our neutron-proton interaction parameter
$v_{\rm np}$ we compare with phenomenological contact interactions
that have been used to model nuclear spectra.  The matrix 
element of the neutron-proton 
interaction is
\be
\langle n_1 p_1 | v | n_2 p_2\rangle = -v_0 I
\ee
where
\be
I = \int d^3 r   \phi_{n_1}^*(\vec r)\phi_{p_1}^*(\vec r)
\phi_{n_2}(\vec r) \phi_{p_2}(\vec r).
\label{eq-I}
\ee
The parameter $v_0$ is the strength of the interaction, typically
expressed in units of MeV fm$^3$.  
Some values
of $v_0$ from the literature are tabulated in Table \ref{v0}.  
\begin{table}[htb] 
\begin{tabular}{|c|c|c|} 
\hline 
Basis of estimate  &  $v_0$ (MeV fm$^3$) &  Citation \\
\hline 
$G$-matrix    & 530  &  \cite{BBB92} \\
$sd$-shell spectra & 490 & \cite{ba88}\\        
$\beta$-decay & 395,320 & \cite{yo13}\\
\hline 
\end{tabular} 
\caption{Estimates of neutron-proton interaction strength.
\label{v0} 
}
\end{table} 
We shall adopt the value $v_0 = 500 $ MeV fm$^3$ for most of the
model calculations.  

\begin{figure}
\includegraphics[width=1.0\columnwidth]{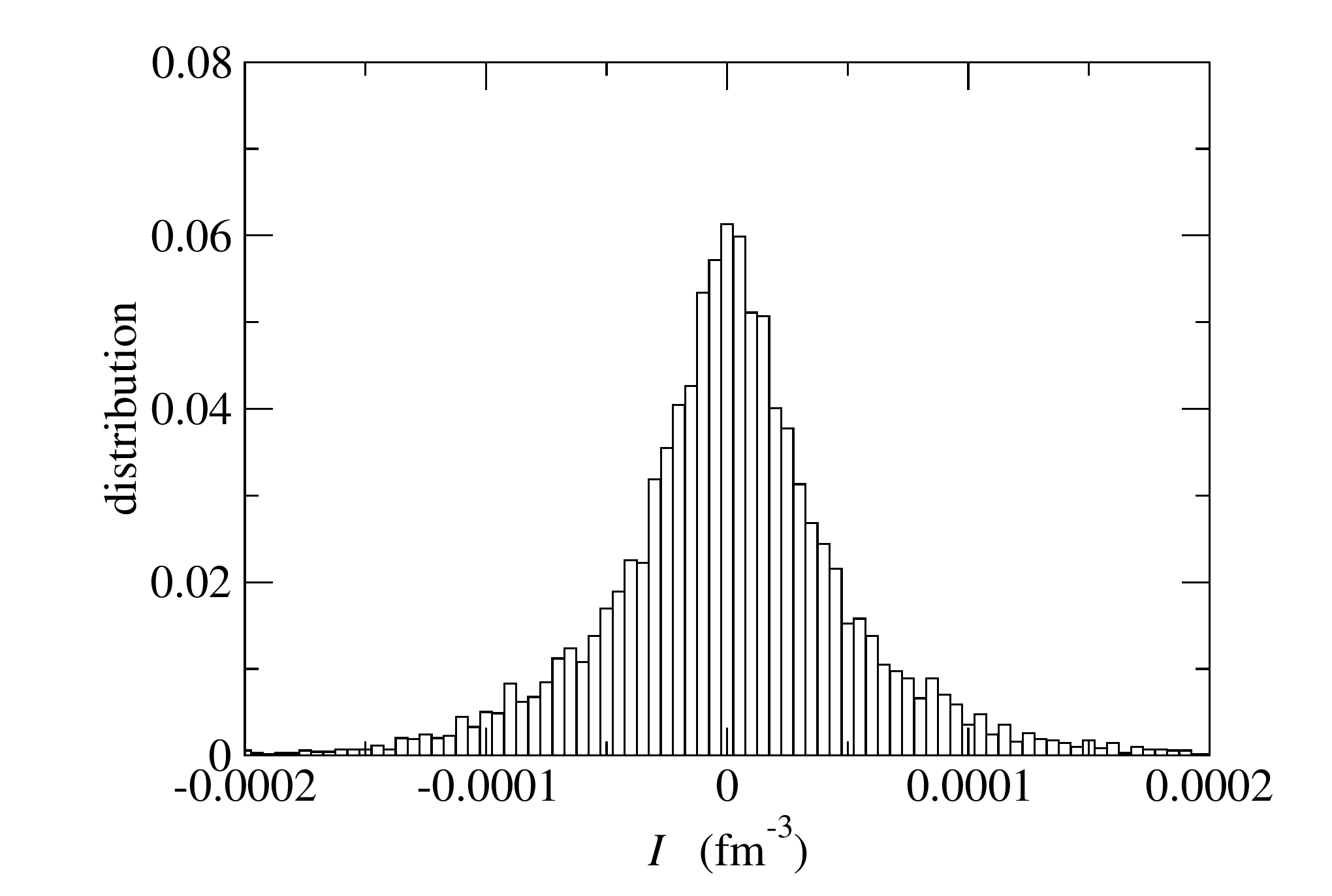}
\caption{Integrals $I$  in Eq. (\ref{eq-I}) of orbitals near
the Fermi energy.  
\label{v_np-hist}
}
\end{figure}

If the wave functions
of the eigenstates approach the compound nucleus limit,
the only characteristic we need to know is its mean-square
average among the active orbitals.  We have calculate the 
integral Eq. (\ref{eq-I}) for all the fully off-diagonal matrices
of the orbitals within 2 MeV of the Fermi energy. 
Fig. \ref{v_np-hist} shows a histogram of their distribution 
 \footnote{If the 
orbitals are restricted only to those in TABLE II, the histogram is more 
structured.}.
The variance of the distribution is 
$\langle I^2\rangle^{1/2} = 5.22\times 10^{-5}$ fm$^{-3}$.  
Combining this with our estimate of $v_0$ we find 
$(\overline{\langle n_1 p_1 |v |n_2 p_2\rangle^2})^{1/2} = 0.025$ MeV.
This implies $v_{\rm np} \sim 0.05 d$ with our estimated single-particle
level density.\\

\noindent{\it Decay widths}
Experimentally in the nucleus \uf~at an energy near $S_n$, the
compound-nucleus gamma decay widths are about $0.040$ eV\cite{ca09}.
Such  widths are smaller than any of the other energy scales in the 
reaction. 
The branching ratio between gamma capture and fission favors fission
by about a factor of 3, but $B_{\rm cap,fiss}$ has strong fluctuations
around that average. There is no direct information about the exit
channel decay widths near the scission point. However, for several
of the examples in the text we have taken $\Gamma_{\rm fis}/\Gamma_{\rm cap}=3$, 
which would give the observed branching in the (unphysical) 
compound nucleus limit.

On the theoretical side, the decay widths of the states in the Hamiltonian enter into
the reaction cross sections into two ways, explicitly as a
factor in Eq. (5) of the text and implicitly in the Green's 
function  $(\tilde H - E)^{-1}$.  If the decay width is smaller than any
of the other internal energy scales, one can neglect its effect on the
Green's function.  It is also the case that the transmission factors
depend strongly on the entrance channel widths, but the branching ratios
are insensitive.  Our reported calculations were carried out in the
limit of small entrance channel widths, but realistic ones derived
from optical model phenomenology can be easily incorporated, as was
done in the MAZAMA code \cite{be17}.\\

\noindent{\bf Compound nucleus limit.}\\
\begin{figure}
\includegraphics[width=\columnwidth]{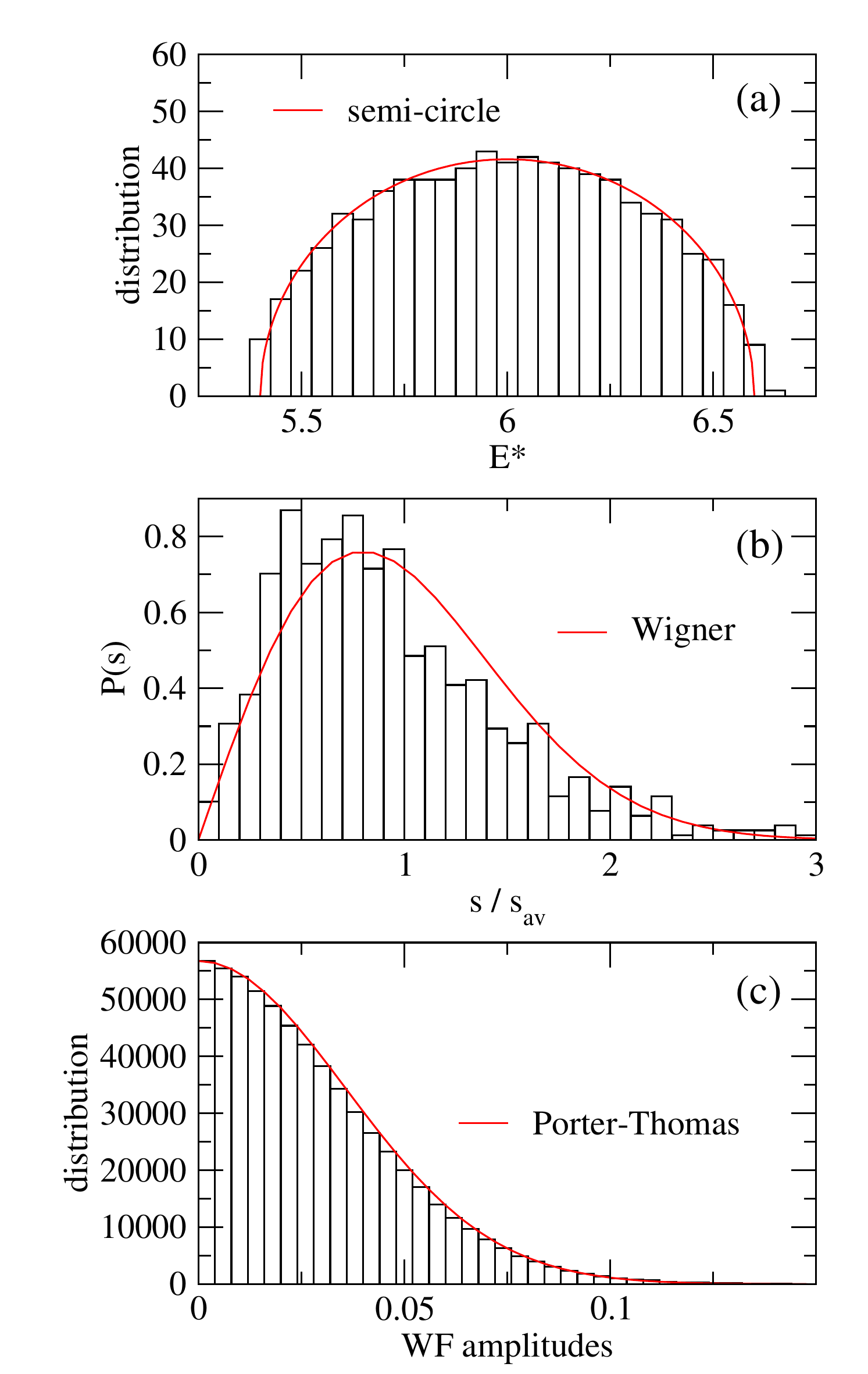}
\caption{(a) Eigenvalue spectrum of the neutron-proton interaction with 
$v_{\rm np}=0.05d$ in the $E^*=6d$ configuration space. 
The red solid curve shows the semi-circle distribution, Eq. (\ref{semi-circle}), 
with $E_0=0.6d$. 
(b) The nearest neighbor level spacing distribution compared with the 
Wigner distribution, Eq. (\ref{wigner}). 
(c) The distribution of wave function amplitudes for 
all the components of 
all the eigenstates within 
the model space. The red solid line shows the expected Gaussian distribution.} 
\label{fig:goe}
\end{figure}
The concept of the compound nucleus is a major
ingredient of the reaction theory for fission of heavy nuclei. 
From the side of theory, the compound nucleus is characterized 
by a set of properties derived from Wigner's random 
matrix model (RME) 
\cite{gu89,wei09,Zurbauer83,drozdz94,matsuo93,matsuo97,matsuo97-2},
or more specifically the Gaussian orthogonal ensemble (GOE). 
In our framework, the RME is most closely approached when
the space is restricted to a single $k$-block.
Since the diagonal elements of the Hamiltonian are all the same, 
the only energy scale relevant to the diagonalization is $v_{\rm np}$, 
the strength of the interaction. 
However, it is far from guaranteed that the model will satisfy the
expected properties because our Hamiltonian is a sparse
matrix, unlike the RME.
Here we examine the properties of the states in the $k=6$ space
to show that indeed the model does approach the RME limit.
The Hamiltonian matrix has a 
dimension of $N_6=784$ and there are 29688 off-diagonal matrix 
elements. To be specific, the interaction strength is taken
as $v_{\rm np} = 0.05d$.

Properties of the RME that are independent of the
basis are the semicircular distribution of eigenvalues and
the repulsion between neighboring eigenvalues. 
The top and the middle panels of Fig. \ref{fig:goe} 
show the eigenvalue distribution and the distribution of the nearest neighbor 
level spacing compared with the semi-circle formula 
\begin{equation}
\rho(E)=\frac{2}{\pi}\,\frac{N_6}{E_0}\,\sqrt{1-\left(\frac{E}{E_0}\right)^2},
\label{semi-circle}
\end{equation}
and the eigenvalue repulsion formula (i.e., the Wigner distribution), 
\begin{equation}
P(x)=\frac{\pi}{2}xe^{-\pi x^2/4}
\label{wigner}
\end{equation}
respectively. 
Here, $E_0$ in Eq. (\ref{semi-circle}) is a parameter characterizing the 
width of the eigenvalue distribution, which we take $E_0=0.6d$, 
and $x$ in Eq. (\ref{wigner}) is defined as 
$x=s/\langle s\rangle$, where $s$ is a nearest neighbor level spacing 
and $\langle s\rangle$ is its average. 
The figure indicates that both properties are reasonably well satisfied. 

For the problem of fission, 
the most important property of the compound nucleus is 
a distribution of decay widths to the individual channels. This 
should obey the Porter-Thomas distribution 
if the configuration amplitudes are Gaussian distributed. 
Taking all the eigenfunctions and
configurations of the model space, 
the bottom panel of Fig. \ref{fig:goe} 
shows the distribution of the wave function amplitudes 
compared with the Gaussian
distribution having the width 
parameter $\sigma=N_6^{-1/2}$. 
As we see, the distribution agrees well with the Gaussian distribution. \\

\noindent{\bf Codes}\\

Codes to compute selected data points in Figs. 1,3,4, and 5 are
provided in the subdirectory  {\tt codes}.  The Hamiltonian
matrices are constructed by Fortran codes {\tt ham*.f}.  The
subsequent analysis is carried out by Python scripts and 
shell scripts.  The Fortran codes have been compiled and tested 
with the gfortran compiler.  The Python scripts are listed below.
They compile the necessary Fortran code to generate the
Hamiltonians in the ensembles and then calculate the observables
according to the formulas in the text.
 \\
Fig. 1, histogram:    fig1.py\\
Fig. 3, branching ratio at   $f_{\rm od} = 0.5$:  fig3.py\\
Fig. 4, branching ratio at $\Gamma_{\rm fis}= 0.1$:\\ 
        fig4.py\\
Fig. 5, $B$ at $N_k = 16$:  fig5.py\\

%